\documentclass[reprint,superscriptaddress,amsmath,amssymb]{revtex4-2}

\usepackage{graphicx}
\usepackage{dcolumn}
\usepackage{bm}
\usepackage{mathrsfs}
\usepackage{comment}
\usepackage[caption=false]{subfig}

\begin{document}

\preprint{APS/123-QED}

\title{Absence of Phase Transition in Random Language Model}

\author{Kai Nakaishi}
\affiliation{
 Graduate School of Arts and Sciences,
 The University of Tokyo,
 Komaba, Meguro-ku, Tokyo 153-8902, Japan
}
\author{Koji Hukushima}
\affiliation{
 Graduate School of Arts and Sciences,
 The University of Tokyo,
 Komaba, Meguro-ku, Tokyo 153-8902, Japan
}
\affiliation{
Komaba Institute for Science, The
University of Tokyo, 3-8-1 Komaba, Meguro-ku, Tokyo 153-8902, Japan
}

\begin{abstract}
The Random Language Model,
proposed as a simple model of human languages,
is defined by the averaged model of a probabilistic context-free grammar.
This grammar expresses
the process of sentence generation
as a tree graph
with nodes having symbols as variables.
Previous studies proposed that
a phase transition,
which can be considered to represent the emergence of order in language,
occurs in the random language model.
We discuss theoretically that the analysis of
the ``order parameter'' introduced in previous studies
can be reduced to solving the maximum eigenvector
of the transition probability matrix determined by a grammar.
This helps analyze the distribution of a quantity determining the behavior of the ``order parameter'' and reveals that no phase transition occurs.
Our results suggest the need to study a more complex model
such as a probabilistic context-sensitive grammar, in order for phase transitions to occur.
\end{abstract}

\maketitle

\section{Introduction}

Chomsky attempted to establish formal models
that represent the universal grammar
underlying all languages
\cite{Chomsky_SS, Chomsky_three}.
One such model is
a context-free grammar (CFG)
\cite{Chomsky_certain},
which has been used beyond the scope of linguistic research
in various fields
such as quasicrystal\cite{Escudero}
and molecular optimization\cite{Kajino}.
To better understand natural languages,
a possible extension of this abstract model
is a probabilistic version,
called probabilistic context-free grammar (PCFG)
\cite{Jelinek_PCFG}, which has been studied
as a model of cosmic inflation
\cite{Harlow},
recurrent neural network
\cite{LinTegmark},
a prior distribution of RNA secondary structures
\cite{Knudsen},
etc.
However, partly because Chomsky strongly denied
the importance of statistical approaches to language,
limited research has been conducted on PCFG
in the context of understanding natural languages.
Several studies \cite{Klein, Noji, Ney} on
grammar induction and parsing based on PCFG,
which are analogies
for how humans acquire and recognize their first language,
focus on specific corpora
and not on the
universal or typical properties of languages.

Recently, DeGiuli, in their study on the typical properties of PCFG, introduced
a model of averaged PCFG,
called the Random Language Model (RLM)\cite{DeGiuli_RLM}.
This model can be viewed as a
statistical-mechanical model of random systems,
which helps derive the free energy formula of the model using theoretical physics methods,
such as the replica method and
Feynman diagrams\cite{DeGiuli_emergence}.
Numerical simulations and statistical-mechanical
analyses of the model
suggest that a phase transition occurs
between ordered and random phases as the model parameters vary,
demonstrating
the emergence of order in language
just as a child initially speaks incoherently, but later learns to speak grammatically correct language.
The presence of the phase transition would implies that the difference between children's incoherent ``languages'' and adults' languages is not quantitative but qualitative.
However,
the previous analyses
do not prove the existence of the phase transition.
We address this issue because it
is a fundamental problem 
whether a phase transition exists in a simple model such as the RLM.

This article is organized as follows: in Sec.~\ref{sec_model}, the models we discuss, i.e., CFG, PCFG, and RLM, are described  step by step,
and the ``order parameter'' introduced in the previous study\cite{DeGiuli_RLM} to detect a phase transition in the RLM is presented. 
Sec.~\ref{sec_analyses} is the main content of this article where we examine the possibility of the phase transition by analyzing the ``order parameter.'' 
We point out that the probability of occurrence of a symbol is an essential quantity characterizing the RLM, and our analyses in terms of the probability reveal that the phase transition cannot exist.
In other words, the RLM only presents a crossover.
We also discuss how the crossover occurs.
In Sec.\ref{sec_conclusion}, we summarize the results and imply the need to study a more complex model with context-sensitiveness.
The appendices provide further details on our analysis and Shannon entropy. 

\section{Models and ``order parameter''}
\label{sec_model}

\subsection{Context-free Grammar}

\begin{figure}[b]
  \includegraphics[width = .95 \columnwidth]{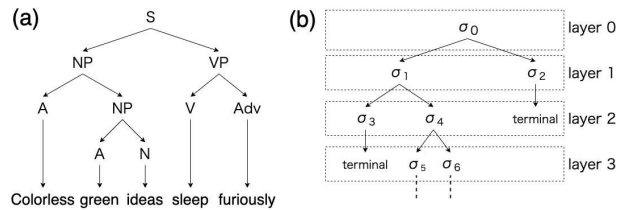}
  \vspace{1ex}
  \caption{
    (a) Example of IC analysis.
    The diagram shows the syntactic structure of
    the sentence ``Colorless green ideas sleep furiously.''
    (b) Layer decomposition of a derivation generated by PCFG.
  }
  \label{fig_IC_layer}
\end{figure}

A conventional approach in linguistics
is to analyze
syntactic structures underlying sentences
in terms of immediate constituent (IC) analyses.
In this framework,
these structures are represented graphically
by tree diagrams as shown in Fig.~\ref{fig_IC_layer}(a)
\cite{Chomsky_SS}.
CFG or type 2 grammar
is a formal grammar which generates sentences
with such tree-like structures
\cite{Chomsky_certain, Chomsky_three}.
The CFG $G$ consists of
a set $\Sigma$ of symbols called a vocabulary or an alphabet,
and a set $R$ of rules.
The vocabulary $\Sigma$
is the union of two disjoint finite sets
$\Sigma_N$ and $\Sigma_T$,
which are sets of nonterminal and terminal symbols,
respectively.
Each rule in $R$ is of the form $A \to \varphi$,
i.e., an instruction to rewrite
a single nonterminal symbol $A$
as a string $\varphi$ in $\Sigma$.
Given the special nonterminal symbol $S$,
called a starting symbol,
one applies rules in $R$ iteratively
until the string comprises solely terminal symbols.
This final string of terminal symbols is defined as a sentence generated by $G$.
Furthermore, the complete process of generating a sentence
is called a derivation.
A derivation is represented as a tree
as in IC analyses,
the leaves of which represent the sentence.
In linguistics,
nonterminal symbols correspond to
constituents or phrases
such as sentences, noun phrases, or verb phrases,
while terminal symbols correspond to words or morphemes,
such as ``ideas'' or ``sleep''.
Although
CFG does not explain all aspects of natural languages,
it reflects the fundamental structure
of a class called kernel
comprising basic sentences
such as active and declarative sentences.
Without the loss of generality,
one can restrict rules to the form
$A \to B C$ and $A \to a$,
where $A$, $B$, and $C$ are nonterminal symbols
and $a$ is a terminal symbol
\cite{hopcroft}.
We shall call them nonterminal rules and terminal rules,
respectively.

\subsection{probabilistic Context-free Grammar}

Sentences we generate depend
not only on the grammar,
but also on the intentions of speakers or writers and
situations where they are generated.
However, for simplicity,
we consider the model in which
sentences are generated randomly
according to given probabilistic weights and CFG,
i.e., PCFG
\cite{Jelinek_PCFG}.
We assign weight $M_{ABC}$ to
each $(A, B, C) \in \Sigma_N^3$,
and $O_{Aa}$ to
each $(A, a) \in \Sigma_N \times \Sigma_T$.
$M_{ABC}$ and $O_{Aa}$ define the probability that
nonterminal symbol $A$ is rewritten by
nonterminal rule $A \to BC$
and terminal rule $A \to a$, respectively.
Notice that $M_{ABC} = 0$ or $O_{Aa} = 0$ in PCFG
corresponds to the situation that
$R$ does not include $A \to BC$ or $A \to a$ in CFG.
The sets $M = \{ M_{ABC} \}$ and $O = \{ O_{Aa} \}$
of weights determine the probability with which
each sentence or derivation is generated.
With fixed $\Sigma$,
each PCFG is only characterized by $M$ and $O$.

According to
\cite{DeGiuli_RLM, DeGiuli_emergence},
we introduce the ``emission probability'' $p$,
which is the probability that a nonterminal symbol
is rewritten as a terminal symbol,
and redefine weights $M_{ABC}$ and $O_{Aa}$ such that
the probability for $A \to BC$ and $A \to a$
are $(1 - p) M_{ABC}$ and $p O_{Aa}$, respectively.
In this setup, the topology $\mathcal{T}$ of a derivation
is independent of the occurrence of symbols
on $\mathcal{T}$, 
and is only controlled by the emission probability $p$.

\subsection{Random Language Model}

We are interested in
typical properties of a general language,
not of a particular language.
To analyze the typical properties,
we consider an averaged model, assuming
that weights $(M, O)$ are generated randomly
according to some distribution $P_{M, O} (M, O)$.
The introduction of the emission probability allows us to divide
$P_{M, O} (M, O)$ into
$P_M ( M )$ and $P_O (O)$.
As these distributions, we choose lognormals:
\begin{equation}
  \begin{split}
    P_M (\tilde{M}_{ABC})
    &\propto \mathrm{e}^{
      - \epsilon_M \ln^2 \tilde{M}_{ABC}
    }, \
    M_{ABC}
    = \frac{
      \tilde{M}_{ABC}
    }{
      \sum_{B, C} \tilde{M}_{ABC}
    }, \\
    P_O (\tilde{O}_{Aa})
    &\propto \mathrm{e}^{
      - \epsilon_O \ln^2 \tilde{O}_{Aa}
    },\
    O_{Aa}
    = \frac{\tilde{O}_{Aa}}{\sum_a \tilde{O}_{Aa}},
  \end{split}
  \label{eq_lognormal}
\end{equation}
where $\epsilon_M$ and $\epsilon_O$ are parameters that characterize the prior distributions. This is the RLM
introduced by DeGiuli\cite{DeGiuli_RLM, DeGiuli_emergence}.

\subsection{``Order Parameter''}
\label{subsec_order}

In the previous 
studies,
to clarify the behavior of the RLM,
the ``order parameter'' $Q_{ABC}$ is introduced as
\begin{align*}
  & Q_{ABC} (M; p) \\
  &\equiv \left\langle
    \frac{1}{ \Omega ( \mathcal{T} ) }
    \sum_{(i, j, k) }
    \delta_{\sigma_i, A} \left(
      N^2
      \delta_{\sigma_j, B}
      \delta_{\sigma_k, C}
      - 1
    \right)
  \right\rangle_{M, p},
\end{align*}
where $\Omega (\mathcal{T})$ is the number of applications
of nonterminal rules in a derivation.
$i$, $j$, and $k$ are the indices of nodes of
given $\mathcal{T}$,
and $\sigma_i$ is a nonterminal symbol on node $i$.
The summation runs over all $(i, j, k)$
such that $\sigma_i \to \sigma_j \sigma_k$
is applied in a derivation.
$\langle \cdots \rangle_{M, p}$ denotes
the average over derivations generated according to
weights $M$ and the emission probability $p$.
Note that we do not need to consider weights $O$
as long as we focus on $Q_{ABC}$
because this parameter depends only on the structure of nonterminal symbols.
This definition is motivated by that of the order parameter
for Potts model\cite{gross_1985_Potts}.

Numerical simulations of this quantity
suggest the existence of a phase transition with a change in the parameter $\epsilon_M$.
The transition point separates
a random phase and an ordered phase.
In the former,
the averaged sum $\left[ \sum_{A,B,C} Q_{ABC}^2 \right]_{\epsilon_M}$
of squared ``order parameters'' is vanishingly small,
where $[\cdots]_{\epsilon_M}$ means the average over weights $M$ according to $\epsilon_M$.
Meanwhile, in the latter,
it takes a finite value.
This may be interpreted as an indication of
the emergence of nontrivial order,
which allows
sentences to communicate information.
However, the origin and characteristics of the singularity associated with a phase transition are not always evident.
Indeed, the previous studies present no concrete evidences
for the phase transition.

\section{Analysis}
\label{sec_analyses}

\subsection{Probability of Occurrence of Symbol}

In this paper, we theoretically revisit the phase transition in the RLM
from a viewpoint different from that of previous studies.
The key point is that the singularity of
``order parameter'' $Q_{ABC}$
is reduced to that of the probability of occurrence
of symbol $A$.
First, we denote this as
\begin{align*}
  \pi_A ( M; p )
  &\equiv \left\langle
    \frac{
      \mbox{$\#$ of symbol $A$ in a derivation}
    }{
      \mbox{the size of a derivation}
    }
  \right\rangle_{M, p},
\end{align*}
where the nonterminal symbols
rewritten by terminal rules
are not counted in both the numerator and the dominator.
The ``order parameter'' $Q_{ABC}$
can be rewritten
as
\begin{align}
  Q_{ABC} (M; p)
  = N^2
  \pi_A
  \left(
    M_{ABC} - \frac{1}{N^2}
  \right).
  \label{eq_Q_ABC_pi}
\end{align}
Because the distribution $P_M$ is an analytic function of $\epsilon_M$,
the behavior of factor
$M_{ABC} - 1 / N^2$ is also analytic.
Thus,
the order parameter $Q_{ABC}$ will not present any singularity
unless the distribution of $\pi_A$
non-analytically changes.

For the analysis of $\pi_A$,
it is useful to decompose a tree into layers
$0,1, 2, \cdots$ from the root to leaves,
as shown in Fig.~\ref{fig_IC_layer}(b).
In addition, we denote the number of nodes
and occurrence of symbol $A$ in layer $d$,
except those turning into terminal symbols in layer $d + 1$,
as $l_d$ and $n_A^{(d)}$, respectively.
In terms of these, $\pi_A$ is represented as
\begin{align}
  \pi_A (M; p)
  &= \lim_{D \to \infty}
  \sum_{d = 0}^D \left\langle
    \frac{
      n^{(d)}_A
    }{
      L^{(D)}
    }
  \right\rangle_{M, p},
  \label{eq_piA_rewritten}
\end{align}
where $L^{(D)} \equiv \sum_{d' = 0}^D l_{d'}$.
Recall that the topology $\mathcal{T}$ of a derivation
is generated independently of
the occurrence of symbols on $\mathcal{T}$.
Therefore, we can average over the occurrence
$\bm{\sigma}^{(D)} = ( \sigma_0, \cdots, \sigma_{L^{(D)} - 1} )$ of symbols up to layer $D$
before the topology $\mathcal{T}$ in the computation of the summand,
represented as
\begin{align*}
  \left\langle
    \frac{
      n^{(d)}_A
    }{
      L^{(D)}
    }
  \right\rangle_{M, p}
  &= \underset{\mathcal{T}}{\mathbb{E}} \left[
    \frac{l_d}{
      L^{(D)}
    } \underset{\bm{\sigma}^{(D)}}{\mathbb{E}} \left[
      \left.
        \frac{n_A^{(d)}}{l_d}
      \right|
      \mathcal{T}
    \right]
  \right] .
\end{align*}
Considering the process of developing the layers,
the transition probability 
that
a symbol $A$ in layer $d$
turns into a symbol $B$ in layer $d + 1$ 
is given by
\begin{align*}
  W_{B A} (M)
  &\equiv \frac{
    \sum_C M_{A B C}
    + \sum_C M_{A C B}
  }{2} \\
  &= \frac{
  \sum_C \tilde{M}_{A B C}
  + \sum_C \tilde{M}_{A C B}
  }{2 \sum_{B, C} \tilde{M}_{A B C}}.
\end{align*}
Using the transition matrix $W$, it turns out
\begin{align*}
  \underset{\bm{\sigma}^{(D)}}{\mathbb{E}} \left[
    \left.
      \frac{\bm{n}^{(d)}}{l_d}
    \right|
    \mathcal{T}
  \right]
  = W^d ( M ) \bm{\pi}^{(0)},
\end{align*}
where $\bm{n}^{(d)}
\equiv ( n_0^{(d)}, \cdots, n_{N - 1}^{(d)} )$
and $\bm{\pi}^{(0)} \equiv ( \pi_0^{(0)}, \cdots, \pi_{N - 1}^{(0)} )$.
Notice that $\bm{\pi}^{(0)}$ is the distribution of starting symbols.
The right hand side of the above equation is independent of the topology $\mathcal{T}$.
Combining these,
the summand of Eq.~(\ref{eq_piA_rewritten}) yields
\begin{align}
  \left\langle
    \frac{
      \bm{n}^{(d)}
    }{
      L^{(D)}
    }
  \right\rangle_{M, p}
  &= f_d^{(D)} (p)
  W^d ( M ) \bm{\pi}^{(0)},
  \label{eq_expected_nd} \\
  f_d^{(D)} (p)
  &\equiv \underset{\mathcal{T}}{\mathbb{E}} \left[
    \frac{l_d}{L^{(D)}}
  \right].
\end{align}
It would be interesting to see that
Eq.~(\ref{eq_piA_rewritten}) is separated into the product form of the dependencies on $p$ and $M$.

\subsection{Case of $p = 0$}

The factor $f_d^{(D)} (p)$ can be expressed in terms of
a moment generating function (see Appendix~\ref{app_generating}).
For general $p$,
it is difficult to write it down explicitly,
but in the special case of $p=0$, the following argument can be proceeded.
In this case, $l_d = 2^d$
and 
$f^{(D)}_d ( 0 ) = 2^d / ( 2^{D + 1} - 1 )$.
Substituting these into
Eq.~(\ref{eq_piA_rewritten}) and (\ref{eq_expected_nd}),
it turns out that $\pi_A = v_A$,
where $\bm{v} \equiv \lim_{d \to \infty} W^d \bm{\pi}^0$
is the steady state of the Markov chain defined by $W$.
Because $W$ is a transition probability matrix,
$\bm{v}$ corresponds to the unique maximum eigenvector of $W$ belonging to eigenvalue 1 with probability $1$
if $\epsilon_M > 0$.
As a consequence, it turns out that for $p = 0$,
it is sufficient to find the largest eigenvector of $W$
to analyze the ``order parameter'' $Q_{ABC}$
of a given grammar $M$.
For $\epsilon_M > 0$,
it is convenient to define a set
$\mathcal{M} ( \pi_A )
\equiv \{ \tilde{M} > 0 \ | v_A ( \tilde{M} ) = \pi_A \}$
for each $\pi_A \in [0, 1]$,
where
$\tilde{M} > 0$ means that $\tilde{M}_{ABC} > 0$ in $\tilde{M} = \{ \tilde{M}_{ABC} \}$ for any $(A, B, C)$.
The probability density of $\pi_A$ is expressed as 
$P ( \pi_A | \epsilon_M, p = 0 )
= \int_{\mathcal{M} ( \pi_A )} \mathrm{d} \tilde{M}
P_M ( \tilde{M} )$.
Because the measure $P_M ( \tilde{M} )$
is an analytic function of $\epsilon_M$
defined by Eq.~(\ref{eq_lognormal}),
$P ( \pi_A | \epsilon_M, p = 0 )$
is also analytic with $\epsilon_M$.
This means that a phase transition
detected by the ``order parameter'' $Q_{ABC}$ cannot 
exist.

\subsection{Case of $0 < p < 1$}

For $0 < p < 1$,
how does the distribution of $\pi_A$ behave
when $\epsilon_M$ changes?
As $p$ increases,
$f^{(D)}_d ( p )$ for small $d$ increases
while that for large $d$ decreases,
and thus $P ( \pi_A )$ gets closer to
$( 1 / N ) \sum_A \delta ( \pi_A - \pi^{(0)}_A )$
since $W^d \bm{\pi}^{(0)}$ for smaller $d$
is closer to $\bm{\pi}^{(0)}$.
It is unlikely that this effect leads to a singularity.
This implies that no phase transition exists
not only for $p = 0$ but also for general $p$.

To confirm the above argument quantitatively,
we measured numerically the Binder parameter of $\pi_A$,
defined as
\begin{align}
  U
  \equiv 1
  - \frac{
    \left[
      \left( \Delta \pi_A \right)^4
    \right]_{\epsilon_M}
  }{
    3 \left[
      \left( \Delta \pi_A \right)^2
    \right]_{\epsilon_M}^2
  },
  \label{eq_Binder}
\end{align}
where
$\Delta \pi_A
\equiv \pi_A - [ \pi_A ]_{\epsilon_M}
= \pi_A - 1 / N$.
This parameter
has been used 
to numerically detect the transition temperature
of first- and second-order transitions in various statistical-mechanical models
\cite{Binder1981, Binder_Landau, Binder_Landau}.
For all previously known cases of phase transitions
detected by this parameter,
a discontinuous jump
from zero implying a Gaussian distribution
to a finite value
determined by a multimodal distribution
is found at the transition temperature
in the thermodynamic limit.

\begin{figure}[b]
  \includegraphics[width = .95 \columnwidth]
  {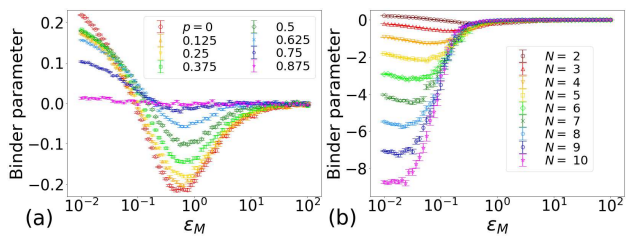}
  \caption{
    (a) $\epsilon_M$ dependence of
    the Binder parameter of $\pi_A$
    for $N=2$ and $p = 0, 0.125, \cdots , 0.875$.
    Results are computed by eigenvalue analysis for $p=0$, and by numerical simulations 
    with a maximum depth $D = 7$ for $0 < p < 1$.
    (b) Binder parameters of $\pi_A$
    for $N = 2, \cdots, 10$ and $p = 0$
    computed through eigenvalue analysis.
    In both (a) and (b),
    error bars are evaluated by the bootstrap method
    with $5 \times 10^2$ bootstrap sets\cite{Young}.
  }
  \label{fig_Binder}
\end{figure}

Fig.~\ref{fig_Binder}~(a) shows the Binder parameters
of $\pi_A$ for $N = 2$
with several values of the emission probability $p$. 
For $p = 0$, we directly calculated $\pi_A$
for the derivations of the maximum depth $D \to \infty$,
i.e., those of infinite sizes,
using the solution of the maximum eigenvector problem
of transition probability matrix $W$.
This is a significant advantage of the fact that $\pi_A = v_A$.
For $0 < p < 1$, we 
measured $\pi_A$ by implementing the RLM.
Because it is impossible to generate
a derivation of infinite size,
we measured $\pi_A$ with the maximum depth $D = 7$ instead,
which was sufficient to approximate
the Binder parameter for $D \to \infty$.
We chose $\pi_A^{(0)} = 1 / N$ for any symbol $A$
as the distribution of starting symbols,
thus the Binder parameter of $\pi^{(0)}_A$ is zero.
The number of grammars $M$ generated for each $\epsilon_M$
was $4\times 10^5$ for both $p = 0$ and $0 < p < 1$.
In addition, $10^3$ derivations were generated
for each grammar $M$ for $0 < p < 1$.
In this figure, the Binder parameter for $p = 0$
is shown to be analytic in $\epsilon_M$,
which is predicted from the above analysis
concluding that the distribution
$P ( \pi_A | \epsilon_M, p = 0 )$ is analytic.
It should be emphasized that
this plot is calculated directly in terms of eigenvalue analysis,
not approximated by generating derivations of finite sizes.
This plot means that the Binder parameter at $p=0$ is truly analytic for an infinite system, not that it appears to be analytic 
because of finite system.
From this figure, we can also see that
the Binder parameter get closer to zero,
with no singularities for any $p$,
as $p$ increases.
This is consistent with the above argument
that there is no phase transition for $0 < p < 1$.

Note that the distribution $P ( \pi_A )$
may have a singularity with respect to $p$ at $p=1/2$,
caused by percolation transition\cite{percolation}.
However, even if this transition exists,
we will be able to explain it as a phenomenon of a tree branching probabilistically, 
and this singularity has no relation
to the weights $M$ which is the essential element of PCFG.
This transition 
would not reflect
any 
nature of PCFG.

\subsection{Mechanism of Crossover}

Even if a quantity does not present a phase transition,
it will be an interesting problem how the quantity crosses over.
The fact that $\pi_A=v_A$ for $p=0$ helps 
understand the behavior of $P (\pi_A)$ at $p = 0$.
In the $\epsilon_M \to \infty$ limit,
$M_{ABC}$ for any $A$, $B$, and $C$
is $1 / N^2$ with probability 1,
leading to $W_{AB} = 1 / N$
for any $A$ and $B$,
and the maximum eigenvector $\bm{v}$
is $( 1 / N, \cdots, 1 / N )$.
This implies $P ( \pi_A ) = \delta ( \pi_A - 1 / N )$.
Meanwhile, for the $\epsilon_M = 0$,
one element of $M_{ABC}$'s is 1
and the other $N^2 - 1$ elements are 0
for each $A$ with probability 1.
Thus, the number of possible $M$
is the same as the choice of
the elements of value 1,
and the numbers of
possible $W$ and the corresponding $\bm{v}$ are finite.
Thus, $P ( \pi_A )$ has multiple delta peaks.
Note that the maximum eigenvector is not unique
and $P ( \pi_A )$ depends on the initial state $\bm{\pi}^0$
in this case.
In the intermediate region $0 < \epsilon_M < \infty$,
$P ( \pi_A )$ is a continuous function of $\pi_A$
because $P_M (M)$ is that of $M$.
If $P (\pi_a)$ turned discontinuously
from single delta peak to multiple ones at finite $\epsilon_M$,
it would be a phase transition similar to
that with one-step replica symmetry-breaking\cite{Binder_Young, beyond}.
However, $P ( \pi_A )$ becomes a delta function
only at $\epsilon_M = 0$ and $\epsilon_M \to \infty$.

\begin{figure}[t]
  \includegraphics[width = .95 \columnwidth]
  {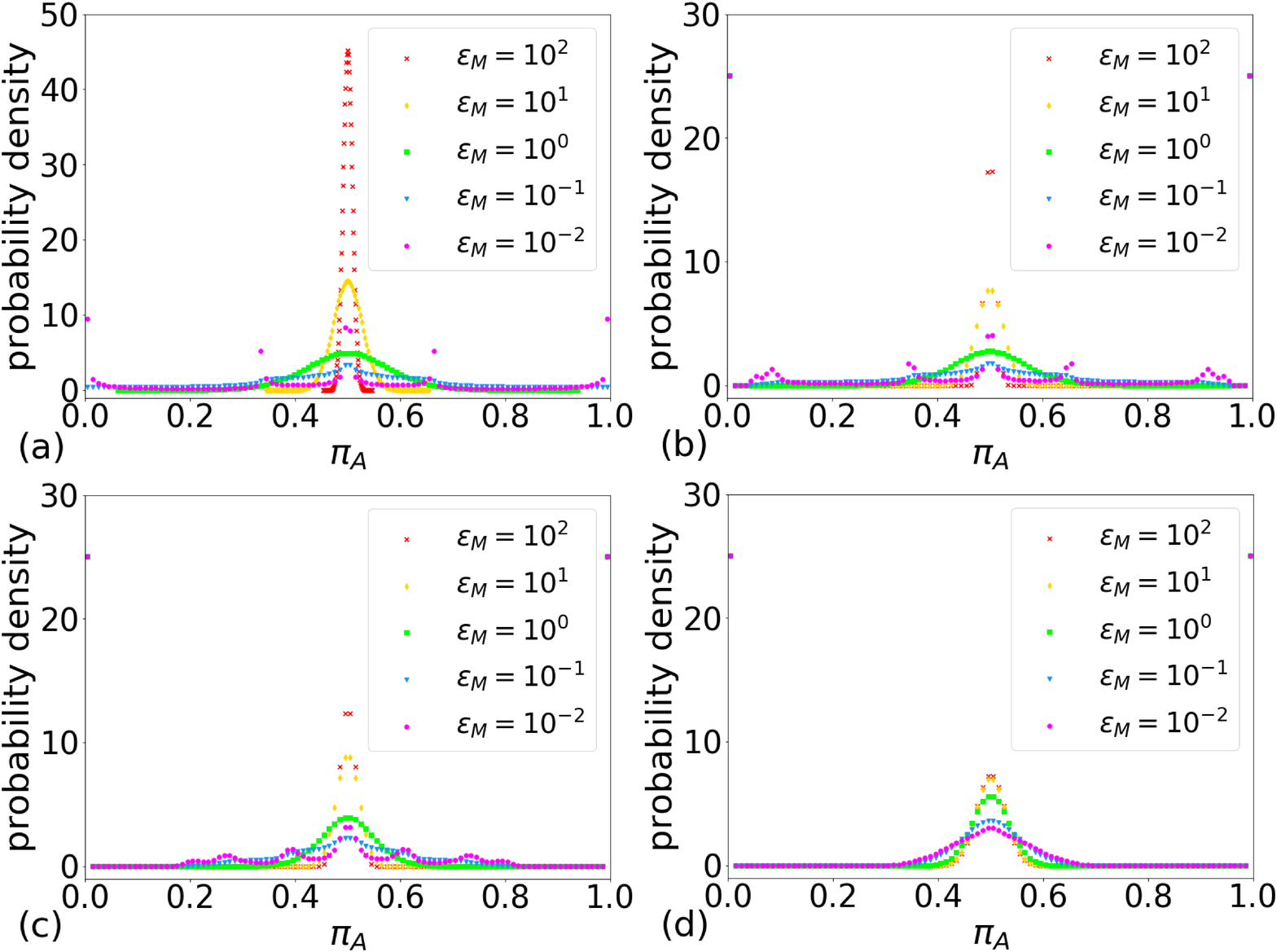}
  \caption{
    Distribution of $\pi_A$ for
    (a) $p = 0$, (b) $p = 0.25$, (c) $p = 0.5$,
    and (d) $p = 0.75$.
    $\pi_A$ for $p = 0$ was computed through
    eigenvalue analysis,
    while that for $0 < p < 1$ was approximated
    implementing the RLM with the fixed maximum depth $D = 7$.
  }
  \label{fig_pdf}
\end{figure}

We actually computed and observed the distribution $P (\pi_A)$ for $N = 2$ and $p = 0, 0.25, 0.5, 0.75$
through numerical simulations.
As in measurement of the Binder parameters,
$\pi_A$ for $p = 0$ was computed through
eigenvalue analysis 
while that for $0 < p < 1$ was approximated
implementing the RLM with the fixed maximum depth $D = 7$.
We assumed $\pi^0_A = 1 / N$
for any nonterminal symbol $A$.
In addition,
we generated $4\times 10^5$ grammars $M$ for each $\epsilon_M$
for both $p = 0$ and $0 < p < 1$
and $10^3$ derivations for each grammar $M$
for $0 < p < 1$.
Fig.~\ref{fig_pdf}(a) shows the plots for $p = 0$.
$P ( \pi_A )$ for $p = 0$ has
a single peak around $1 / N$,
meaning that symbols occur according to
almost uniform distribution under many grammars.
Meanwhile, for small $\epsilon_M$,
$P ( \pi_A )$ is multimodal,
because various grammars are generated
and their distributions $\bm{\pi}$ of symbols
are different depending on grammars.
However, we emphasize that
$P ( \pi_A )$ is not a delta function
for finite $\epsilon_M$
as discussed in the preceding paragraph.
This figure also show that the dip of the Binder parameter in Fig.~\ref{fig_Binder} occurs around the point at which $P (\pi_A)$ turns from single peak to multiple peaks.
From Fig.~\ref{fig_pdf}(b), (c), and (d),
it can be seen that $P ( \pi_A )$ becomes closer to
Gaussian centered at $\pi_A = 1 / 2 = \pi_A^{(0)}$
as $p$ increases.
This is also consistent with the above discussion.

\subsection{The Limit $N \to \infty$}

\begin{figure}[t]
  \includegraphics[width = .95 \columnwidth]
  {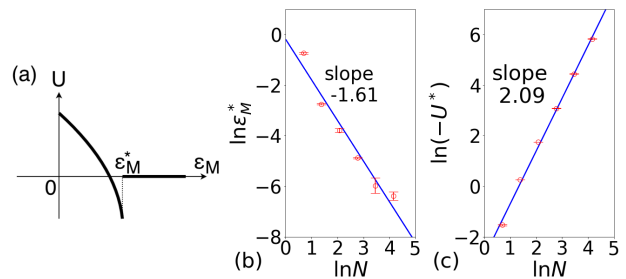}
  \caption{
    (a) A possible behavior of Binder parameter with singularity
    in the limit $N \to \infty$.
    (b) Local minimum point $\epsilon_M^*$
    of the Binder parameter as a function of $N$
    in log-log plot.
    (c) Absolute value of the local minimum $U^*$
    of the Binder parameter
    as a function of $N$ in log-log plot.
    In both (b) and (c),
    the straight lines represent
    linear regression using the least squares method,
    and the error bars are evaluated by the bootstrap method
    with $5 \times 10^2$ bootstrap sets\cite{Young}.
  }
  \label{fig_LM_Bin}
\end{figure}

It is revealed that
there is no phase transition for finite $N$.
We now consider the possibility of
existence of phase transition in the limit $N \to \infty$.
If we agree with the concept of the
so-called {\it infinite use of finite means},
the essential characteristic of language
is that it can express infinite meanings
with finite symbols\cite{Chomsky_SS,yang}.
From this viewpoint, it is not clear what property of language this limit explains.
Still, this limit is interesting as a physical model,
since PCFG is used as a model for various phenomena
other than language.
As we have shown,
it is sufficient to deal with the case with $p = 0$
for the investigation of whether a phase transition exists.
Fig.~\ref{fig_Binder}(b) shows
the Binder parameters of $\pi_A$
for $N = 2, \cdots, 10$ and $p = 0$.
For each $\epsilon_M$,
$4\times 10^5$ grammars for $N = 2, \cdots, 7$
and $1.6\times 10^6$ grammars for $N = 8, 9, 10$
were generated.
As seen in Fig.~\ref{fig_Binder}(b), 
the dip of the Binder parameter gets deeper as $N$ increases.
In the limit $N \to \infty$, this point
might become singular at finite $\epsilon_M$
shown in Fig.~\ref{fig_LM_Bin}(a),
which can be a phase transition
as seen in a three-dimensional Heisenberg spin glass
\cite{Ima-Kawa_02}.

To study whether this phenomenon
occurs,
we observed numerically the dependence of
the local minimum point of the Binder parameter on $N$
by generating $4\times 10^5$ grammars for each $N$ and $\epsilon_M$.
As Fig.~\ref{fig_LM_Bin}(b) and (c) show, the results
suggest
that
$\epsilon_M^*$ converges to zero algebraically in $N$
and $U^*$ also diverges to $-\infty$ algebraically.
Thus, no evidences imply the existence of
singularity as in Fig.~\ref{fig_LM_Bin}(a)
in the limit $N \to \infty$.

\section{Conclusion}
\label{sec_conclusion}

Our analysis for finite $N$ holds
even if the weights $M$ are not generated from a lognormal distribution.
As long as the distribution is an analytic function
of $\epsilon_M$,
the analyticity of $P ( \pi_A | \epsilon_M, p)$
can be shown theoretically
in the same way.
Our numerical analysis for infinite $N$
also suggests that no phase transition exists
even in the limit $N \to \infty$.
This conclusion might be different
if the distribution of $M$ is not the lognormal distribution.
For example, 
if
$M_{ABC}$'s are not i.i.d., but dependent on each other,
then
$P ( \pi_A | \epsilon_M, p )$ might have a singularity
with respect to $\epsilon_M$ in the limit $N \to \infty$.
In Ref.~\cite{DeGiuli_RLM}, DeGiuli also
claimed that Shannon entropy has a singularity.
However, an argument similar to that for the ``order parameter'' $Q_{ABC}$ can be developed to show
the analyticity of Shannon entropy.
The detailed explanation is presented in Appendix~\ref{app_Shannon}.

In terms of both the ``order parameter'' $Q_{ABC}$
and Shannon entropy,
the RLM does not have a singularity.
This may imply that children's languages and adults' languages are different only quantitatively, not qualitatively.
As seen from the overall analysis,
the most fundamental reason for the absence of singularity
relies on the fact 
that the distribution of a symbol on a node
depends only on the weights $M$
and the distribution on its parent node.
To see a nontrivial physical phenomenon
such as a phase transition,
we might need to consider more complex models
in which the distribution on a node
is determined by more factors.
One possible model is
probabilistic context-sensitive grammar (PCSG),
that is, an extension of context-sensitive grammar (CSG)
by assigning probabilistic weights to the rules.
CSG is one level higher
than CFG in the Chomsky hierarchy\cite{Chomsky_certain},
where a resulting string in each rule
depends not on a single symbol,
but on a substring.
Thanks to this property,
the behavior of the distribution of symbols
can no longer be computed in the same manner as in PCFG.

We conclude that,
in the RLM,
the model for the typical evaluation of PCFG,
there does not exist a phase transition
as has been suggested,
because the PCFG and the RLM are too simple.
This fact may imply that language acquisition of children is a continuous process.
The absence of the nontrivial phenomenon, i.e., a phase transition, does not deny
the significance of the study of
the probabilistic extension of grammars
in the Chomsky hierarchy
as an approach for natural language
from mathematical sciences.
The typical properties of PCSG
should be analyzed in the future to search for
the nontrivial physical phenomena
that are different from those in the RLM.

\begin{acknowledgments}
  We would like to thank
  K.~Kaneko, A.~Ikeda, and A.~Morihata
  for useful discussions.
  This work was partly supported by the World-leading Innovative Graduate Study Program for Advanced Basic Science Course at the University of Tokyo.
\end{acknowledgments}

\appendix

\section{Moment generating function for computing $f^{(D)}_d (p)$}
\label{app_generating}

The function $f^{(D)}_d (p)$ is defined by
\begin{align*}
  f_d^{(D)} (p)
  &\equiv \underset{l_0, \cdots, l_D}{\mathbb{E}} \left[
    \frac{l_d}{
      \sum_{d' = 0}^D l_{d'}
    }
  \right],
\end{align*}
where $l_d$ is the number of nonterminal symbols
to which nonterminal rules apply in layer $d$.
This function can be computed
by differentiating and integrating
the following moment generating function:
\begin{align}
  \mathscr{M}
  \equiv \underset{l_0, \cdots, l_D}{\mathbb{E}} \left[
    \mathrm{e}^{
      h_{\mathrm{all}} \sum_{d = 0}^D l_d
      + \sum_{d = 0}^D h_d l_d
    }
  \right] .
  \label{eq_generating}
\end{align}
Indeed, the relation between $f^{(D)}_d$
and $\mathscr{M}$ is as follows:
\begin{align*}
  f^{(D)}_d (p)
  &= \left.
    \frac{\mathrm{d}}{\mathrm{d} h_d}
    \int_{- \infty}^0 \mathrm{d} h_{ \mathrm{all} }
    \mathscr{M}
  \right|_{h_0 = \cdots = h_D = 0} .
\end{align*}

Because $l_{d + 1}$ only depends on $l_d$,
we can take the expectation in the right-hand side of
Eq.~(\ref{eq_generating}) one by one
from $l_D$ to $l_0$.
First, the expectation over $l_D$ is taken,
we have
\begin{align*}
  &\underset{l_D}{\mathbb{E}} \left[
    \left.
      \mathrm{e}^{ ( h_{\mathrm{all}} + h_D ) l_D }
    \right|
    l_{D - 1}
  \right] \\
  &= \sum_{m = 0 }^{ 2 l_{D - 1} } \begin{pmatrix}
    2 l_{D - 1} \\ m
  \end{pmatrix}
  p^m ( 1 - p )^{ 2 l_{D - 1} - m }
  \mathrm{e}^{
    ( h_{\mathrm{all}} + h_D )
    (2 l_{D - 1} - m)
  } \\
  &= \left\{
    p + (1 - p) \mathrm{e}^{ h_{\mathrm{all}} + h_D }
  \right\}^{ 2 l_{D - 1} },
\end{align*}
where $m$ is the number of nonterminal symbols in layer $D$
that change to terminal symbols in the next layer.
Thus, it yields
\begin{align*}
  \mathscr{M}
  &= \underset{l_0, \cdots, l_{D - 1}}{\mathbb{E}} \left[
    \mathrm{e}^{
      \sum_{d = 0}^{D - 2}
      ( h_{\mathrm{all}} + h_d ) l_d
      + \left(
        h_{\mathrm{all}} + h_{D - 1} + g_{D - 1}
      \right) l_{D - 1}
    }
  \right],
\end{align*}
where
\begin{align}
  g_{D - 1}
  \equiv 2 \ln \left\{
    p + ( 1 - p ) \mathrm{e}^{ h_{\mathrm{all}} + h_D }
  \right\} .
  \label{eq_g_D-1}
\end{align}
By repeating the same calculation
for $l_{D - 1}, \cdots, l_0$ sequentially,
we can finally obtain
\begin{align*}
  \mathscr{M}
  = ( 1 - p ) \mathrm{e}^{
    h_{\mathrm{all}} + h_0 + g_0
  } + p,
\end{align*}
where $g_0$ is defined recursively by Eq.~(\ref{eq_g_D-1}) and
\begin{align*}
  g_{d - 1}
  &\equiv 2 \ln \left\{
    p + (1 - p) \mathrm{e}^{
      h_{\mathrm{all}} + h_d + g_d
    }
  \right\}
\end{align*}
for $d = D - 1, \cdots, 1$.

\section{Analyticity of Shannon Entropy}
\label{app_Shannon}

In \cite{DeGiuli_RLM}, DeGiuli also focused on
Shannon entropy of a sequence
$\bm{\sigma} = ( \sigma_{i_1}, \cdots, \sigma_{i_m} )$
of symbols on any fixed nodes $i_1, \cdots, i_m$,
which we denote as
\begin{align}
  H (\bm{\sigma} | M, O, p)
  \equiv \left\langle
    \ln \frac{1}{
      P \left(
        \bm{\sigma} | M, O
      \right)
    }
  \right\rangle_{M, O, p}.
  \label{eq_Shannon_entropy}
\end{align}
With numerical simulations,
DeGiuli suggested 
that this quantity has a singularity
with $\epsilon_M$ at the same point
where the ``order parameter'' $Q_{ABC}$ does,
that is, the averaged Shannon entropy
$[ H ( \bm{\sigma} | M, O, p ) ]_{\epsilon_M}$
is equivalent to that expected
from strings generated uniformly at random
above the transition point
and substantially smaller than it below the point.
We consider whether this expectation is relevant or not.

From Eq.~(\ref{eq_Shannon_entropy}),
Shannon entropy $H$ will not have a singularity
if the joint probability
$P ( \bm{\sigma} | M, O )$ is analytic.
To investigate the joint probability,
we introduce the conditional probability
$G^{(i, j)}_{AB}
\equiv P ( \sigma_i=A | \sigma_j=B, M)$
that the symbol on node $i$ is $A$,
on the condition that the symbol on the ancestor $j$
of node $i$ is $B$.
The joint probability $P ( \bm{\sigma} | M, O )$
can be represented using weights $M$,
the distribution $\bm{\pi}^{(0)}$ of a starting symbol,
and the conditional probability
$G^{(i, j)}_{AB}$.
Consider the case where the number of nodes is $m = 2$,
i.e., $\bm{\sigma} = ( \sigma_{i_1}, \sigma_{i_2} )$.
For any fixed $i_1$ and $i_2$,
the lowest node of the common ancestors of them
uniquely exists,
and we denote it as $j_0$.
The node $j_0$ has the two children,
one of which is the ancestor of node $i_1$,
and the other of which is that of node $i_2$,
denoted $j_1$ and $j_2$, respectively.
The joint probability
$P ( \sigma_{i_1}, \sigma_{i_2} | M )$
is represented as
\begin{align}
  &P ( \sigma_{i_1} = A_1, \sigma_{i_2} =  A_2 | M )
  \nonumber \\
  &= \sum_{B_1, B_2 C, D} G^{(i_1, j_1)}_{A_1 B}
  G^{(i_2, j_2)}_{A_2 B}
  M_{C B_1 B_2}
  G^{(j_0, 0)}_{C D}
  \pi^0_D.
  \label{eq_joint_probability}
\end{align}
The joint probability $P ( \bm{\sigma} | M, 0 )$
for $m \geq 3$ is represented with an argument similar to that given above. 

The matrix $G^{(i, j)}_{AB}$
can be written as a product of
transition probability matrices.
We label the indices of nodes
between $i$ and $j$ as
$i = k_0, k_1, \cdots, k_{l - 1}$ and $k_l = j$
such that
node $k_{p}$ is the parent of node $k_{p + 1}$
for $p = 0, \cdots, l - 1$.
Next, we define matrices $U^{(L)}$ and $U^{(R)}$ by
\begin{align*}
  U^{(L)}_{BA}
  &\equiv \sum_{C} M_{ABC} ,
  \hspace{2em}
  U^{(R)}_{BA}
  \equiv \sum_{C} M_{ACB},
\end{align*}
respectively.
Using these notations,
the matrix $G^{(i, j)}$ is
\begin{align}
  G^{(i, j)}
  = U^{(\alpha_{l - 1})} \cdots U^{( \alpha_0 )},
  \label{eq_Gij}
\end{align}
where $\alpha_p$ is $L$
if node $k_{p + 1}$ is the left child of node $k_p$,
while $\alpha_p$ is $R$
if node $k_{p + 1}$ is the right child.

Combining Eq.~(\ref{eq_joint_probability})
and (\ref{eq_Gij}),
we can compute the joint probability
$P ( \bm{\sigma} | M, O )$ as a product of
$\bm{\pi}^{(0)}$, $M$, $U^{(L)}$, and $U^{(R)}$.
Thus, if the distance between root node
and a node in $\bm{\sigma}$ is finite,
this joint probability is analytic
for any analytic distribution of $M$.
If the distance is infinite,
the joint probability is mainly determined
by a kind of steady state of the sequence of
transition probability matrices
in the way similar to that in the case with
the ``order parameter'' $Q_{ABC}$.
This implies that Shannon entropy $H$ also
does not have a singularity with $\epsilon_M$.

\bibliography{2021_RLM_2}

\end{document}